# Preliminary Design of a General Electronics Platform for Accelerator Facilities

Jinfu Zhu, Hongli Ding, Haokui Li, Qiaoye Ran, Xiwen Dai, Wei Li, Jiawei Han, Yue Li, Zhiyuan Zhang, Weixin Qiu, Weiqing Zhang

*Abstract*–Many accelerators require considerable electronic systems for tests, verification, and operation. In Shenzhen Superconducting Soft X-ray Free Electron Laser ($S^3$FEL), to meet the early tests and verification of various systems, save development expenses, and improve the reusability of hardware, firmware, and software systems, we have considered the needs of each system and preliminarily designed a general electronics platform based on MicroTCA.4. The Advanced Mezzanine Card (AMC) will place an FPGA Mezzanine Card (FMC) that supports 500 MSPS to 2 GSPS ADC/DAC. We will design two FMC cards on the Rear Transition Module (RTM), which can be used for analog signal conditioning and waveform digitization by 10 MSPS to 250 MSPS ADC/DAC or motor control. The commercial MCH, CPU, power module, and MTCA crate are deployed. This platform can also be applied to other accelerator facilities.

## I. Introduction

IN many accelerator facilities, such as the Free Electron Laser and Synchrotron Radiation facilities [1-4], considerable electronic systems are required for tests, verification, and operation. In Shenzhen Superconducting Soft X-ray Free Electron Laser ($S^3$FEL) [5], for example, we classify electronic systems into accelerator and beamline categories. The accelerator category includes laser systems, microwave systems, beam measurement systems, magnetic power supply systems, etc., while the beamline category includes optical and diagnostic systems, experimental station systems, etc. The requirements of these systems include analog or RF signal conditioning, waveform digitization, microwave control, motor control, etc.

MTCA (Micro Telecommunications Computing Architecture) constitutes a suite of modular and openly standardized specifications designed to facilitate the construction of high-performance computer systems that leverage switched fabric communication [6]. This standard is particularly suited for developing compact rack-mounted systems catering to various applications, including telecommunications, industrial automation, and scientific facilities [7-10].

This paper will first show the hardware architecture and modular boards we designed based MTCA.4 standard. Then, we will introduce various applications for accelerator and beamline categories. Finally, we will propose the next work.

## II. Introduction to the Hardware Architecture

Figure 1 shows the basic structure of the MTCA.4 systems. It has 12 AMC slots. The backplane allows point-to-point communication among AMCs. Management is via IPMB (Intelligent Platform Management Bus). The redundant power module and MCH are designed. The commercial MCH, CPU, power module, and MTCA crate are deployed in our preliminary design.

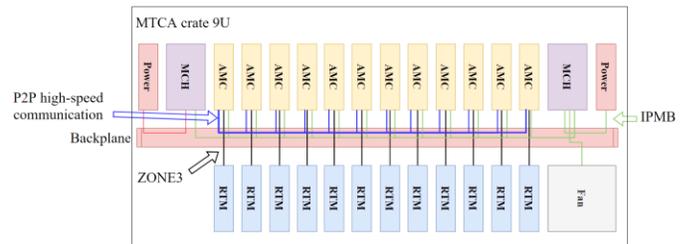

Fig. 1. The diagram of the general electronics platform.

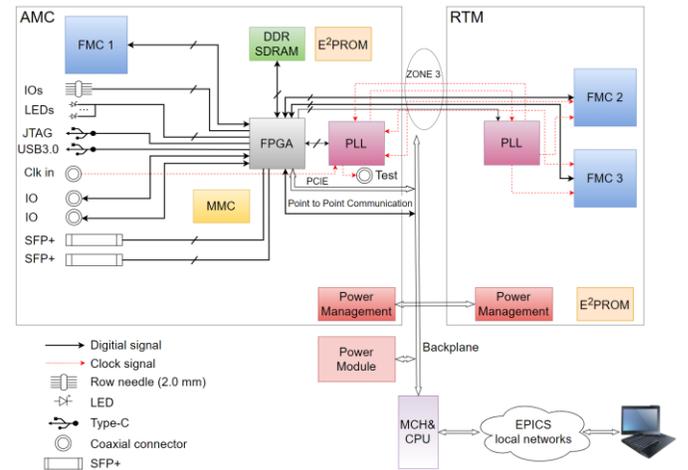

Fig. 2. The diagram of the AMC and RTM boards.

As shown in Fig. 2, we will describe the structure of the AMC and RTM boards. IO pins are used for slow control, and LEDs are for display and monitoring. We will design JTAG on the front panel to make it easy to download FPGA programs. The USB 3.0 interface can conveniently store the

Manuscript received May xx, 2024; revised May xx, 2024. This work was supported by the Shenzhen Science and Technology Program (Grant No. RCBS20221008093247072).

The corresponding authors are Jinfu Zhu and Hongli Ding. Jinfu Zhu is with the Institute of Advanced Science Facilities, Shenzhen, Guangdong, 518107, P.R. China (e-mail: zhujinfu@mail.iasf.ac.cn). Hongli Ding is with the Institute of Advanced Science Facilities, Shenzhen, Guangdong, 518107, P.R. China, and also with the Dalian Institute of Chemical Physics, Chinese Academy of Sciences, Dalian, Liaoning, 116023, P.R. China (e-mail: dinghongli@dicp.ac.cn).

data on a local PC. Three coaxial connectors are connected with FPGA for input/output clocks or controls, and one is connected with PLL for input synchronous clock. Two SFP+ are designed to connect with the timing system and remote data transmission or control, respectively. The AMC board is equipped with an MTCA-based module Management Controller (MMC). DDR SDRAMs are used for data buffering and power management circuits for multiple power sources. Clocks, digital signals, and MMC control ports between RTM and AMC boards are connected via ZONE3 based on D1.0 standard. PC achieves slow control through the EPICS control system network.

## III. APPLICATIONS

AMC and RTM boards support three standard FMC boards. We plan to design four kinds of FMC boards (FMC-A, FMC-B, FMC-C, FMC-D) for early tests and verification of various systems. As shown in Fig. 3 and Fig. 4.

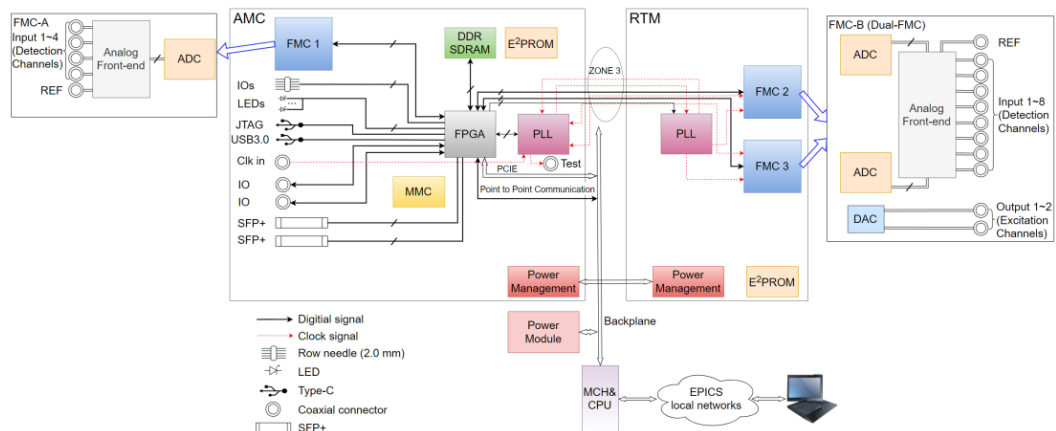

Fig. 3. Applications on GMDs, ICTs, BPMs, Klystron high-voltage or Magnetic power monitors, etc.

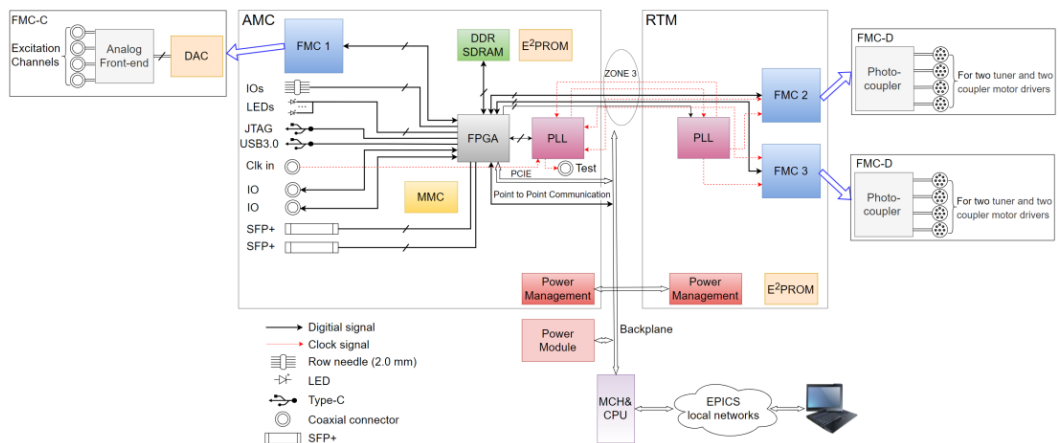

Fig. 4. Applications on motor and piezo control in superconducting cavity.

The FMC-A on the AMC board can support a 500 MSPS to 2 GSPS ADC/DAC for digitization or excitation of the fast pulse in Gas Monitor Detectors (GMDs), Integral Charge Transformers (ICTs), etc., and the RF signal in Beam Position Monitors (BPMs).

The FMC-B (with the dual-FMC area) on the RTM board can realize analog signal conditioning and waveform digitization by 10 MSPS to 250 MSPS ADC/DAC for Klystron high-voltage or magnetic power monitors.

We design motor and piezo control in FMC-C and FMC-D for tuning control in the superconducting cavity. In the S$^3$FEL superconducting cavity tuning system, a 9-cell superconducting cavity requires one piezo control, one tuning motor, and one coupler motor control.

In Figure 4, a pair of AMC and RTM boards support the operation of four superconducting cavities. We use four slow-speed DACs on FMC-C to drive four piezo amplifiers. A single tuning and two coupling motors are connected through two 8-core aviation plug controls.

## IV. NEXT WORK

This paper shows the preliminary design of a general electronics platform for accelerator facilities. This platform can also be applied to other accelerator facilities. We will deploy specific design and testing work in the future.

## REFERENCES

[1] Lee, Sojeong, Kim, Hwan, S., Han, & Jang-Hui, et al. PAL-XFEL cavity beam position monitor pick-up design and beam test[J]. Nuclear

Instruments and Methods in Physics Research, Section A. Accelerators, Spectrometers, Detectors and Associated Equipment, 2016, 827:107-117.
[2] Coughlan J, Cook S, Day C, et al. The data acquisition card for the large pixel detector at the European-XFEL. Journal of Instrumentation, 2011, 6(12), C12057.
[3] Tanaka T, Kato M, Saito N, et al. Absolute laser-intensity measurement and online monitor calibration using a calorimeter at a soft X-ray free-electron laser beamline in SACLA[J]. Nuclear Instruments & Methods in Physics Research, 2018, 894(21):107-110.
[4] Liu, D. K. , Yu, L. Y. , Yin, C. X. , Liu, D. K. , Yu, L. Y. , & Yin, C. X. Digital phase control system for SSRF LINAC[J]. Nuclear Science & Techniques, 2007, 19(3):129-133.
[5] Wang, X., Zeng, L., Shao, J., Liang, Y., Yi, H., Yu, Y., ... & Yang, X. Physical design for shenzhen superconducting soft x-ray free-electron laser (S$^3$FEL)[C]. In Proceedings of International Particle Accelerator Conference, 2023.
[6] https://www.picmg.org/openstandards/microtca/
[7] H Hassanzadegan，A Jansson，C Thomas，A Young，M Werner. System Overview and Current Status of the ESS Beam Position Monitors[C]. In Proceedings of International Particle Accelerator Conference, 2014.
[8] Motuk, E., Postranecky, M., Warren, M., & Wing, M. Experiences with the MTCA. 4 Solution for the EUXFEL Clock and Control System[C]. IEEE-NPSS Real Time Conference (pp. 1-6), 2012.
[9] Branlard, J., Ayvazyan, G., Ayvazyan, V., Grecki, M., Hoffmann, M., Jeżyński, T., ... & Wierba, W. MTCA. 4 LLRF system for the European XFEL[C]. In Proceedings of the 20th international conference mixed design of integrated circuits and systems-MIXDES (pp. 109-112), 2023.